\documentclass[a4paper,11pt]{article}
\usepackage{pos}
\usepackage{braket}
\usepackage{comment}

\title{The emergence of expanding space-time in a novel large-$N$ limit of the Lorentzian type IIB matrix model}
\ShortTitle{The emergence of expanding space-time in the Lorentzian type IIB matrix model}

\author*[a]{Mitsuaki Hirasawa}
\author[b]{Konstantinos N. Anagnostopoulos}
\author[c]{Takehiro Azuma}
\author[d]{Kohta Hatakeyama}
\author[d, e]{Jun Nishimura}
\author[b]{Stratos Kovalkov Papadoudis}
\author[f]{Asato Tsuchiya}

\affiliation[a]{Sezione di Milano Bicocca, Istituto Nazionale di Fisica Nucleare,\\ Piazza della Scienza, 3, I-20126 Milano, Italy}

\affiliation[b]{Physics Department, School of Applied Mathematical and Physical Sciences, National Technical University,\\ Zografou Campus, GR-15780 Athens, Greece}
\affiliation[c]{Institute for Fundamental Sciences, Setsunan University,\\ 17-8 Ikeda Nakamachi, Neyagawa, Osaka, 572-8508, Japan}
\affiliation[d]{KEK Theory Center, High Energy Accelerator Research Organization,\\ 1-1 Oho, Tsukuba, Ibaraki 305-0801, Japan}
\affiliation[e]{Graduate University for Advanced Studies (SOKENDAI),\\ 1-1 Oho, Tsukuba, Ibaraki 305-0801, Japan}
\affiliation[f]{Department of Physics, Shizuoka University,\\ 836 Ohya, Suruga-ku, Shizuoka 422-8529, Japan}

\emailAdd{Mitsuaki.Hirasawa@mib.infn.it}
\emailAdd{konstant@mail.ntua.gr}
\emailAdd{azuma@mpg.setsunan.ac.jp}
\emailAdd{khat@post.kek.jp}
\emailAdd{jnishi@post.kek.jp}
\emailAdd{sp10018@central.ntua.gr}
\emailAdd{tsuchiya.asato@shizuoka.ac.jp}

\abstract{The Lorentzian type IIB matrix model is a promising candidate for a non-perturbative formulation of superstring theory. However, it was found recently that a Euclidean space-time appears in the conventional large-$N$ limit. In this work, we add a Lorentz invariant mass term to the original model and consider a limit, in which the coefficient of the mass term vanishes at large $N$. By performing complex Langevin simulations to overcome the sign problem, we observe the emergence of expanding space-time with the Lorentzian signature.}

\FullConference{%
  The 39th International Symposium on Lattice Field Theory (Lattice2022),\\
  8-13 August, 2022 \\
  Bonn, Germany 
}

\begin{document}
\begin{flushright}
KEK-TH-2485
\end{flushright}
\maketitle

\section{Introduction}
\vspace{-0.25em}
Superstring theory is a promising candidate for the quantum gravity.
The theory is consistently defined in 10-dimensional space-time, although we have observed that our universe is 4-dimensional space-time.
The compactification is a mechanism to describe how to effectively realize 4-dimensional space-times at a low-energy regime in superstring theory.
In the mechanism, there are innumerable number of possible ways to compactify the extra 6-dimensional space perturbatively. 
Therefore it is difficult to choose a unique vacuum that corresponds to our universe perturbatively.

The type IIB matrix model \cite{Ishibashi:1996xs} is a promising candidate for a non-perturbative formulation of superstring theory.
This model is formally obtained by the dimensional reduction \cite{Eguchi:1982nm} of $\mathcal{N}=1$ supersymmetric SU($N$) Yang-Mills theory from 10D to 0D.
The model consists of $N\times N$ Hermitian matrices.
Space-time does not exist a priori, but it emerges from the matrix degrees of freedom.

The Euclidean version of the model was studied analytically using the Gaussian expansion method in \cite{Nishimura:2001sx,Kawai:2002jk,Aoyama:2006rk,Nishimura:2011xy}, and the results predict a spontaneous symmetry breaking (SSB) of the spatial symmetry from SO(10) to SO(3). 
This prediction was confirmed by the first principles calculations using the complex Langevin method (CLM) to avoid a notorious sign problem in \cite{Anagnostopoulos:2013xga,Anagnostopoulos:2015gua,Anagnostopoulos:2017gos,Anagnostopoulos:2020xai}.
On the other hand, the Lorentzian version of the model was studied numerically using an approximation to avoid the sign problem.
The results show an emergence of (3+1)-dimensional space-time \cite{Kim:2011cr, Ito:2013qga, Ito:2013ywa, Ito:2015mxa, Ito:2015mem}, however, the structure of the space is not a continuous one due to the approximation \cite{Aoki:2019tby}.

Quite recently, we studied a fermion quenched version of the Lorentzian type IIB matrix model without the approximation using the CLM \cite{Nishimura:2019qal, Hirasawa:2021xeh}.
We found that the Lorentzian model is equivalent to the Euclidean one under the Wick rotation as it is \cite{Hatakeyama:2021ake}.
In this work, to obtain results inequivalent to the Euclidean one, we use a Lorentz invariant ``mass'' term.
In addition, we study the SUSY model and add a fermionic mass term to avoid the singular drift problem in the CLM \cite{Ito:2016efb}.
We found that the SSB of SO(9) spatial symmetry occurs, however, a 1-dimensional space expands exponentially when the mass of the fermions is large.
We expect that a 3-dimensional expanding space appears when the fermionic mass term is sufficiently small.

\vspace{-0.5em}
\section{Definition of the type IIB matrix model}
\vspace{-0.25em}
The partition function of the Lorentzian type IIB matrix model is written as
\begin{equation}
    Z=\int dA d\Psi d\bar{\Psi}\ e^{i\left( S_{\rm b} + S_{\rm f} \right)},
\end{equation}
\begin{equation}
    S_{\rm b}=-\frac{N}{4} {\rm Tr}\left\{ -2[A_0, A_i]^2 + [A_i,A_j]^2 \right\},
\end{equation}
\begin{equation}
    S_{\rm f}=-\frac{N}{2} {\rm Tr}\left\{ \bar{\Psi}_\alpha (C\Gamma^\mu)_{\alpha\beta}[A_\mu,\Psi_{\beta}] \right\},
\end{equation}
where $A_\mu$ and $\Psi_\alpha$ are $N\times N$ Hermitian matrices, $\mu$ runs from 0 to 9, and $\alpha$ runs from 1 to 16.
$\Gamma^\mu$ are 10-dimensional gamma matrices after the Weyl projection, and $\mathcal{C}$ is a charge conjugate matrix.
This model has $\mathcal{N}=2$ supersymmetry (SUSY), and the SUSY algebra realizes translations by a shift of the matrices $A_\mu \to A_\mu + \alpha_\mu I$, where $I$ is the unit matrix.
Therefore, the eigenvalues of the matrices $A_\mu$ can be interpreted as the space-time coordinates.
After the integration over the fermionic matrices $\Psi_\alpha$, we obtain the following partition function:
\begin{equation}
    Z=\int dA\ e^{iS_{\rm b}} {\rm Pf}{\mathcal M}(A_0,A_i),
\end{equation}
where $\mathcal{M}$ is the Dirac operator, and ``Pf'' stands for Pfaffian.

\begin{figure}
\centering
\includegraphics[scale=0.33]{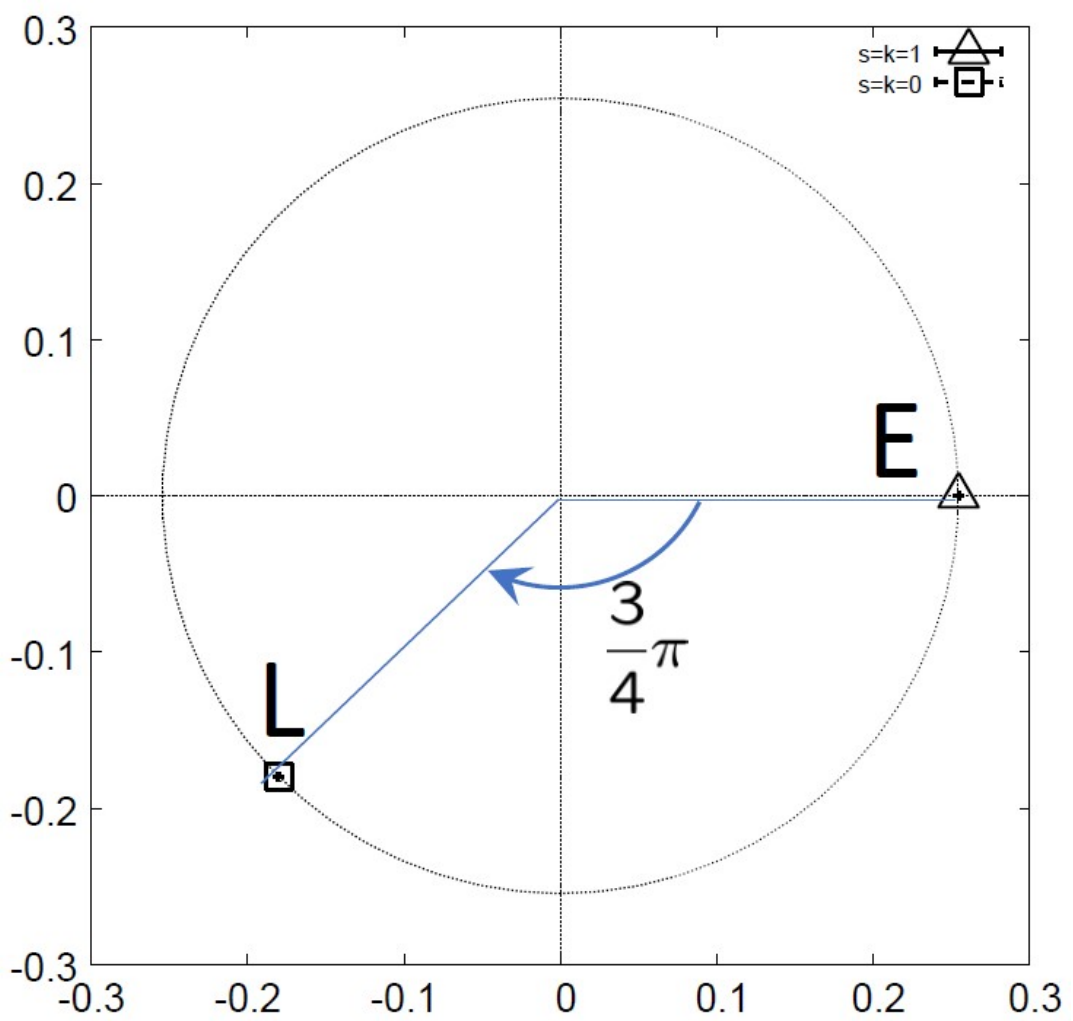}
\includegraphics[scale=0.33]{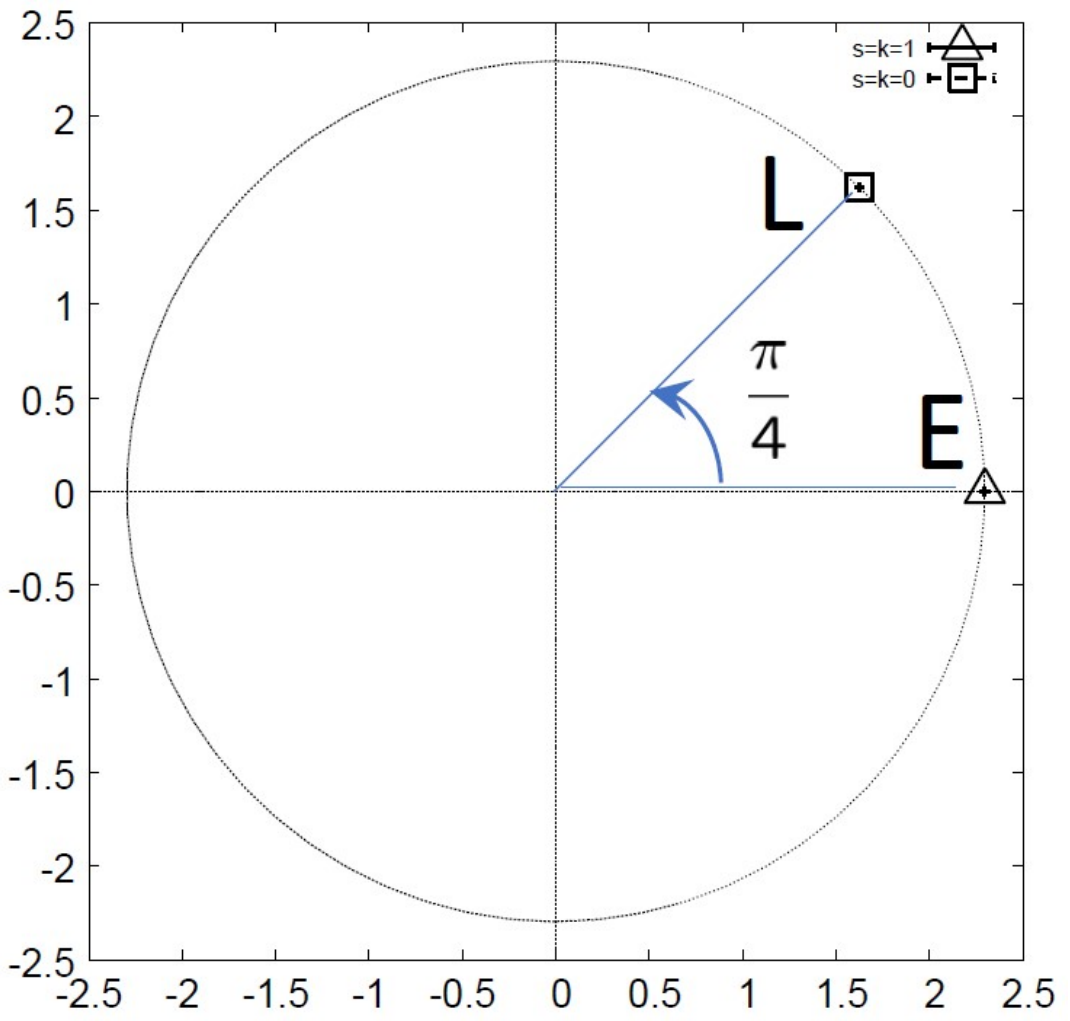}
\caption{${\rm Tr}(A_0)^2$ (Left) and ${\rm Tr}(A_i)^2$ (Right) obtained by simulations of a fermion quenched model. ``E'' and ``L'' correspond to the Euclidean and Lorentzian versions of the model.}
\label{fig:equive_Euc_Lor}
\vspace{-1em}
\end{figure}

We perform a Wick rotation defined by
\begin{equation}
    S_{\rm b} \to \tilde{S}_{\rm b} = N\ e^{i\frac{\pi}{2} u}\ {\rm Tr} \left\{ \frac{1}{2}e^{-i\pi u}[\tilde{A}_0, \tilde{A}_i]^2 - \frac{1}{4}[\tilde{A}_i,\tilde{A}_j]^2 \right\},
\end{equation}
\begin{equation}
    \mathcal{M}(A_0,A_i) \to \mathcal{M}(e^{-i\frac{\pi}{2}u}A_0,A_i),
\end{equation}
where $u=0$ and $u=1$ correspond to the Lorentzian and Euclidean model respectively, and we omit an irrelevant overall phase factor for $\mathcal{M}$ because it can be absorbed by a redefinition of the fermionic matrices.
This Wick rotation is equivalent to the following contour deformation:
\begin{equation}
\begin{split}
    A_0 &\to e^{-i\frac{\pi}{2}u}e^{i\frac{\pi}{8}u}\tilde{A_0} = e^{-i\frac{3}{8}\pi u}\tilde{A_0},\\
    A_i &\to e^{i\frac{\pi}{8}u}\tilde{A_i}.
\end{split}
\label{eq:cont_deform}
\end{equation}
The Cauchy's theorem says that $\braket{\mathcal{O}(e^{-i\frac{3}{8}\pi u}\tilde{A_0}, e^{i\frac{\pi}{8}u}\tilde{A_i})}_u$ is independent of $u$.
This fact means that the Lorentzian version of this model is equivalent to the Euclidean one under the above Wick rotation.
We confirmed this fact using the complex Langevin simulation of the fermion quenched model in \cite{Hatakeyama:2021ake} (See Fig.\ref{fig:equive_Euc_Lor}).

In order to obtain a large-$N$ limit which is inequivalent to the Euclidean model, we add a Lorentz invariant ``mass'' term to the action given by
\begin{equation}
    S_\gamma = -\frac{1}{2} N \gamma  {\rm Tr}(A_\mu)^2 = \frac{1}{2} N \gamma \left\{ {\rm Tr}(A_0)^2 - {\rm Tr}(A_i)^2 \right\},
\end{equation}
where $\gamma$ is the Lorentz invariant ``mass'' parameter.
This term was introduced for the first time in \cite{Steinacker:2017vqw} as an IR regulator, and has been studied at the perturbative level in \cite{Steinacker:2017bhb, Sperling:2018xrm, Sperling:2019xar, Steinacker:2019dii, Steinacker:2019awe, Steinacker:2019fcb, Hatakeyama:2019jyw, Steinacker:2020xph, Fredenhagen:2021bnw, Steinacker:2021yxt, Asano:2021phy, Karczmarek:2022ejn, Battista:2022hqn}.
Especially in \cite{Hatakeyama:2019jyw}, the authors found
that for $\gamma > 0$, typical classical solutions of fixed dimensionality describe the emergence of an expanding (3+1)-dimensional space-time.
In this work, we perform a first principles calculation of this model to study whether such a vacuum appears dynamically.

\vspace{-0.5em}
\section{Complex Langevin method}
\vspace{-0.25em}
In this section, we explain the complex Langevin method \cite{Parisi:1983mgm,Klauder:1983sp}, which is a method used to overcome the notorious sign problem.
Since the method is an extended version of the (real) Langevin method, we start by a brief review of it.

Here we consider a lattice theory that is described by a real dynamical variable $\phi_n$, where $n$ is a label for the lattice points, and we assume that the action $S(\phi)$ takes real values.
The partition function is given by
\begin{equation}
    Z=\int \prod_n d\phi_n e^{-S\left(\phi_n\right)}.
\end{equation}
In the real Langevin method, the dynamical variable is a solution of the Langevin equation
\begin{equation}
    \frac{d\phi_n}{dt_{\rm L}} = -\frac{\partial S}{\partial \phi_n} + \eta_n(t_{\rm L}),
\end{equation}
where $\partial S/\partial \phi_n$ is called the drift term, $t_{\rm L}$ is a fictitious time, the so-called Langevin time, and $\eta_n(t_{\rm L})$ is a real Gaussian noise with zero mean, and variance $\sigma^2=2$.
By solving the Fokker-Planck equation, one can confirm that the equilibrium probability distribution for $\phi$ is proportional to $e^{-S(\phi)}$.

If the action takes a complex value, then also the drift term takes complex values, and the real variable $\phi_n$ must be complexified.
We denote the complexified variable by $\varphi_n$, and it is a solution of the complex Langevin equation as
\begin{equation}
    \frac{d\varphi_n}{dt_{\rm L}} = -\frac{\partial S}{\partial \varphi_n} + \eta_n(t_{\rm L}),
\end{equation}
where $\eta_n(t_{\rm L})$ is a real Gaussian noise with zero mean and variance $\sigma^2=2$.
It is known that the complex Langevin method does not always yield correct results.
A criterion for the correct convergence was discovered in \cite{Nagata:2016vkn}.
Its implementation requires to check that the drift histogram falls off exponentially or faster with the magnitude of the drift term, which is easy to implement in the simulations.

We apply the CLM to the Lorentzian type IIB matrix model.
In the simulations we fix a gauge in which $A_0$ is diagonal
\begin{equation}
    A_0 = {\rm diag}(\alpha_1, \alpha_2, \dots, \alpha_N), \hspace{2em} \alpha_1 \le \alpha_2 \le \dots \le \alpha_N.
\label{eq:daiag_a0}
\end{equation}
In order to realize this ordering, we use a change of variables suggested in \cite{Nishimura:2019qal}
\begin{equation}
\alpha_1 = 0, \hspace{2em} {\rm and} \hspace{2em}\alpha_i = \sum_{a=1}^{i-1} e^{\tau_a} \hspace{1em}(2\le i \le N),
\end{equation}
where the $\tau_a$ are new variables, which we treat as dynamical variables.
It is obvious that the ordering (\ref{eq:daiag_a0}) is automatically realized.

Then $\tau_a$ and $A_i$ are complexified, and we generate configurations using the complex Langevin equations
\begin{equation}
\begin{split}
\frac{d\tau_a}{dt_{\rm L}} &= -\frac{\partial S}{\partial \tau_a} + \eta_{a}(t_{\rm L}), \\
\frac{d(A_i)_{ab}}{dt_{\rm L}} &= -\frac{\partial S}{\partial (A_i)_{ba}} + (\eta_i)_{ab}(t_{\rm L}),
\end{split}
\end{equation}
where $\eta_{a}(t_{\rm L})$ is the Gaussian noise, and the matrices $(\eta_a)_{ab}(t_{\rm L})$ are Hermitian matrices whose elements are generated using Gaussian noise.
The drift terms $\frac{\partial S}{\partial \tau_a}$ and $\frac{\partial S}{\partial (A_i)_{ba}}$ are computed for real variables $\tau_a$ and Hermitian matrices $(A_i)_{ab}$, and then we complexify $\tau_a$ and $(A_i)_{ab}$, thereby doing an analytical continuation to preserve holomorphicity.
For all our results we use the above mentioned criterion to check the correct convergence of the CLM.

When there are fermionic degrees of freedom, near-zero eigenvalues cause large drifts, which cause the singular drift problem.
This problem is one of the causes of the wrong convergence of the CLM.
To avoid this problem, we introduce a fermionic mass term given by
\begin{equation}
    S_{m_{\rm f}} = iNm_{\rm f} {\rm Tr}[\bar{\Psi}_\alpha (\Gamma_7\Gamma_8^\dagger\Gamma_9)_{\alpha\beta}\Psi_\beta].
\end{equation}
This fermionic mass term has been used successfully in the simulations of the Euclidean model \cite{Anagnostopoulos:2017gos,Anagnostopoulos:2020xai} \footnote{
The authors in \cite{Kumar:2022giw} have proposed an alternative fermionic mass term that preserves the supersymmetry.
}, and the original model is recovered after carefully taking the $m_{\rm f}\to 0$ limit.

We perform the following technique to stabilize the complex Langevin simulation by replacing
\begin{equation}
    A_i \to \frac{1}{1+\epsilon} \left( A_i + \epsilon A_i^\dagger \right)
\end{equation}
after every update.
This procedure is justifiable when the spatial matrices are nearly Hermitian.
A similar procedure, which is called the dynamical stabilization, has been used in the lattice QCD \cite{Attanasio:2018rtq}.
In this work, we choose $\epsilon=0.01$.

\vspace{-0.5em}
\section{Results}
\begin{figure}
    \centering
    \includegraphics[width=0.40\hsize]{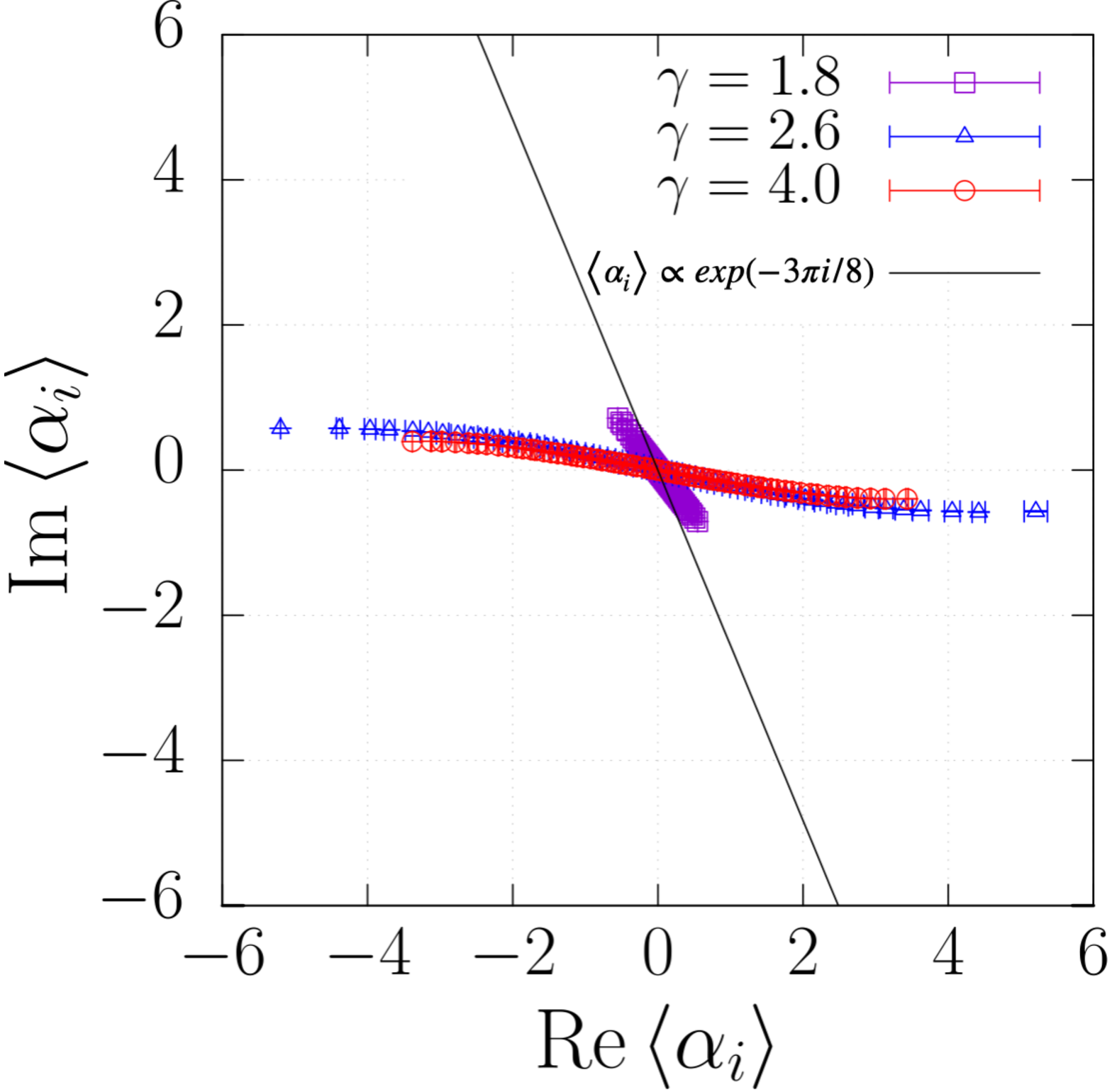}
    \caption{Distributions of the $\alpha_i$ for the Lorentzian model for various values of $\gamma$ at $m_{\rm f}=10$. When the distribution of the $\alpha_i$ approaches the black line, the model is equivalent to the Euclidean one under a contour deformation.}
    \label{fig:phase_structure}
    \vspace{-1em}
\end{figure}

\vspace{-0.25em}
In this paper, all results which we show below are obtained for $N=64$.
In Fig.\ref{fig:phase_structure}, we plot the eigenvalues of $A_0$ on the complex plane for various values of $\gamma$.
If $\gamma$ is equal to, or larger than, 2.6, the eigenvalues are almost real.
On the other hand, if $\gamma$ is equal to, or smaller than, 1.8, the results are equivalent to the Euclidean model under a contour deformation.
We call the former phase the real time phase, and the latter phase the Euclidean phase.
We expect that there is a phase transition for $1.8 \le \gamma \le 2.6$.

\begin{figure}
    \centering
    \includegraphics[width=0.40\hsize]{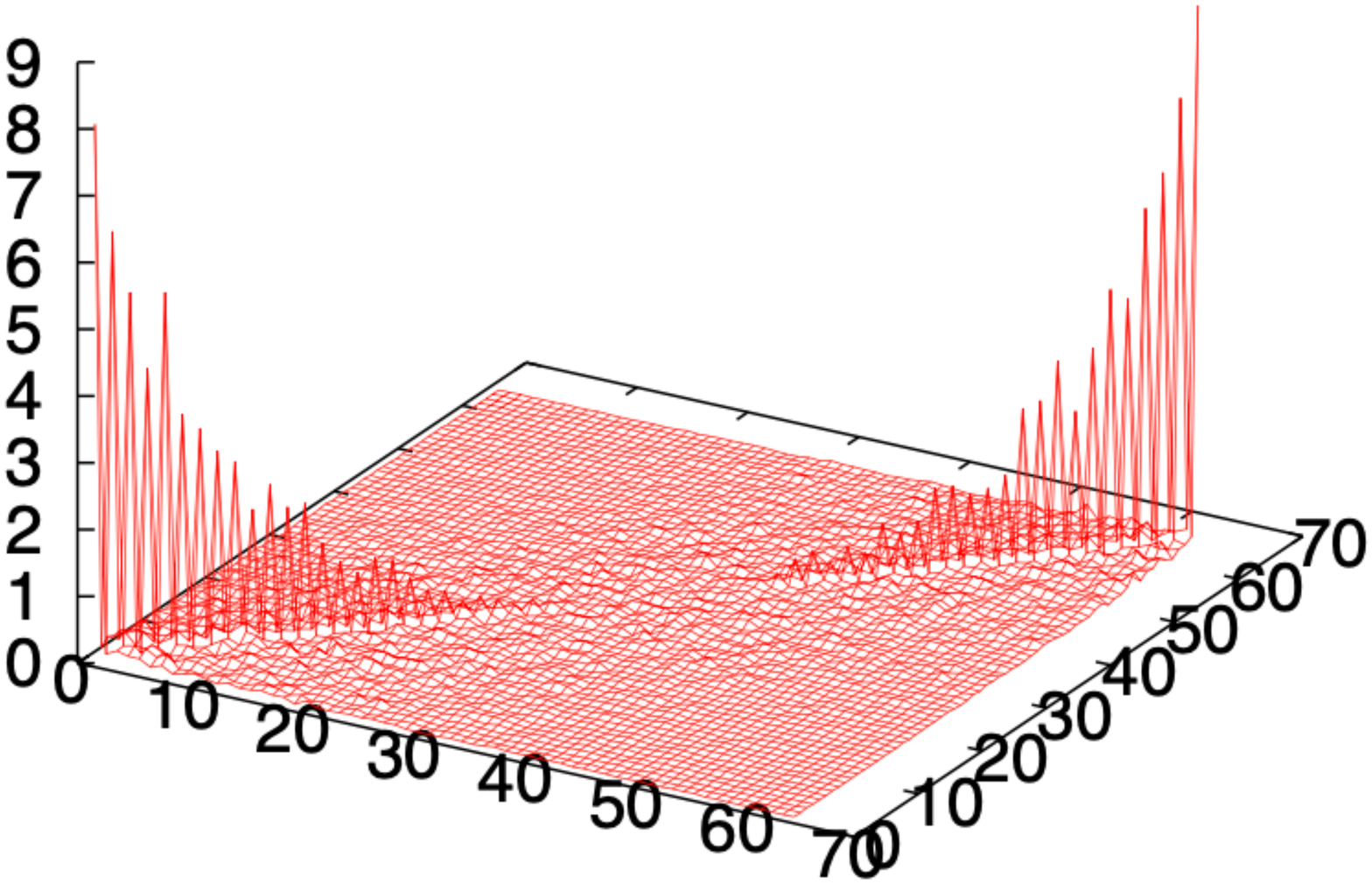}
    \caption{${\mathcal A}_{pq}$ against $p$ ($x$-axis) and $q$ ($y$-axis) at $\gamma=2.6$ and $m_{\rm f}=10$.}
    \label{fig:band_diagonal}
    \vspace{-1em}
\end{figure}

In order to see how the emergence of space from the spatial matrices, we calculate the following observable:
\begin{equation}
    {\mathcal A}_{pq} = \frac{1}{9}\sum_{i=1}^9 |(A_i)_{pq}|^2.
\end{equation}
In Fig.\ref{fig:band_diagonal}, we plot this observables against $p$ and $q$.
This figure shows that the ${\mathcal A}_{pq}$ becomes small as $|p-q|$ increases.
Thus, only the elements near the diagonal have important information.
We call this matrix structure the band-diagonal structure, and $n$ denotes its bandwidth.
In the rest of the paper, we choose $n=12$.

Using the bandwidth $n$, we define time by
\begin{equation}
    t_a = \sum_{i=1}^{a}|\bar{\alpha}_{i}- \bar{\alpha}_{i-1}|,
\end{equation}
where
\begin{equation}
    \bar{\alpha}_i=\frac{1}{n}\sum_{\nu=0}^{n-1} \alpha_{i+\nu}.
\end{equation}
Then we define the $n \times n$ block matrices $\bar{A}_i(t_a)$ by
\begin{equation}
    (\bar{A}_i)_{kl}(t_a) = (A_i)_{(k+a-1)(l+a-1)}.
\end{equation}
These block matrices represent the state of the universe at $t_a$.
In the following, we omit the index $a$, and we denote time by $t$.

\begin{figure}
    \centering
    \includegraphics[width=0.40\hsize]{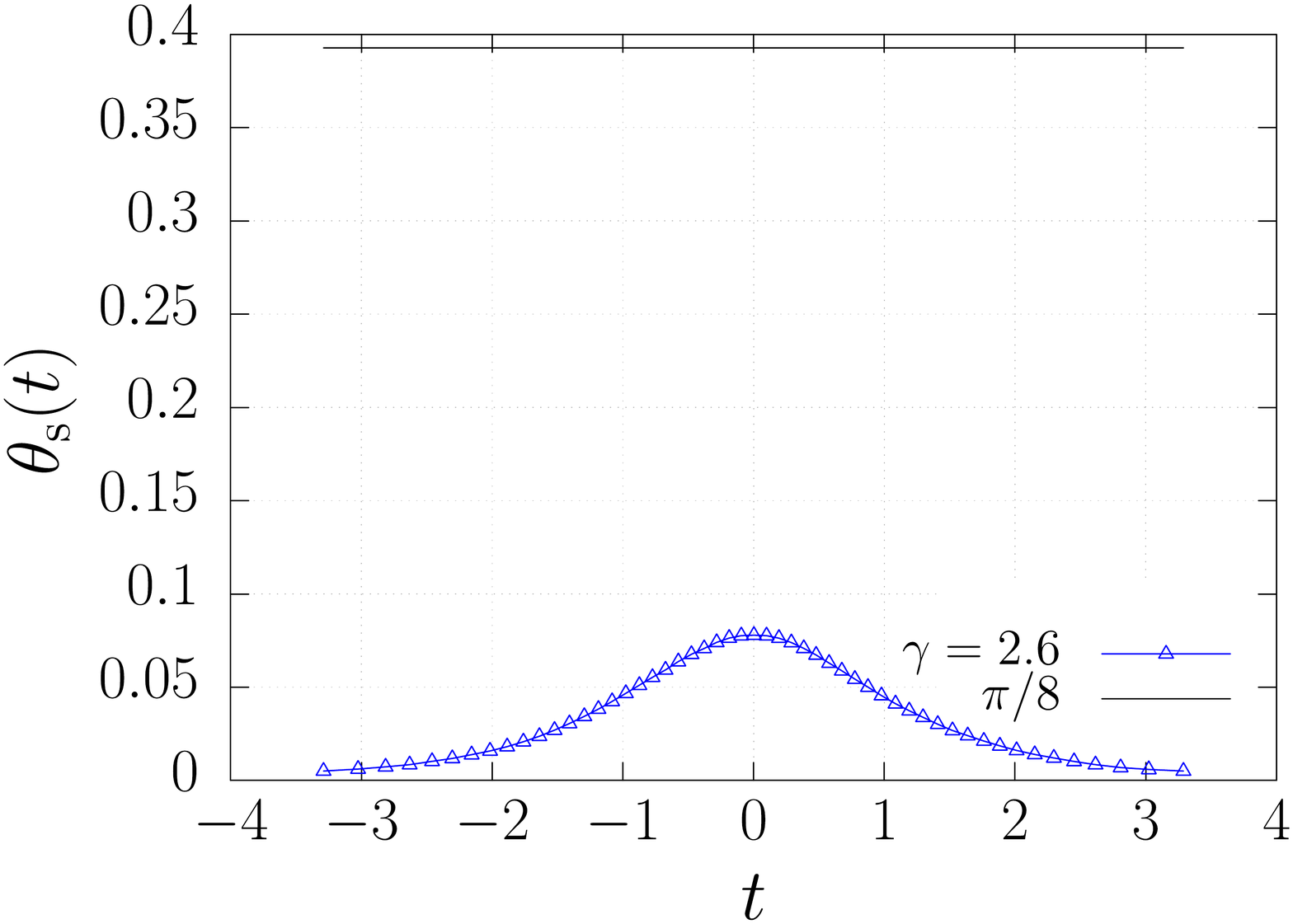}
    \caption{$\theta_{\rm s}(t)$ for $\gamma=2.6$ and $m_{\rm f}=10$. The black line corresponds to the Euclidean model.}
    \label{fig:theta_spatial}
    \vspace{-1em}
\end{figure}

To study whether space is also real in the real time phase, we define a phase $\theta_{\rm s}(t)$ for the spatial matrices by
\begin{equation}
    {\rm tr}(\bar{A}_i(t))^2 = e^{2i\theta_{\rm s}(t)} |{\rm tr}(\bar{A}_i(t))^2|.
\end{equation}
This phase becomes 0 when space is real, $\pi/8$ when the model is equivalent to the Euclidean one.
In Fig.\ref{fig:theta_spatial}, the $\theta_{\rm s}(t)$ is plotted against time.
This figure shows that space becomes real at late times.

To study the SSB of the SO(9) rotational symmetry, we define the ``momentum of inertia tensor'' 
\begin{equation}
    T_{ij}(t) = \frac{1}{n}{\rm tr}\left( X_i(t)X_j(t) \right),
\end{equation}
where
\begin{equation}
    X_i(t) = \frac{1}{2} \left( \bar{A}_i(t) + \bar{A}_i^\dagger(t) \right).
\end{equation}
This Hermitianization is justifiable at late times because the spatial matrices become Hermitian there as we saw in Fig.\ref{fig:theta_spatial}.

\begin{figure}
    \centering
    \includegraphics[width=0.33\hsize]{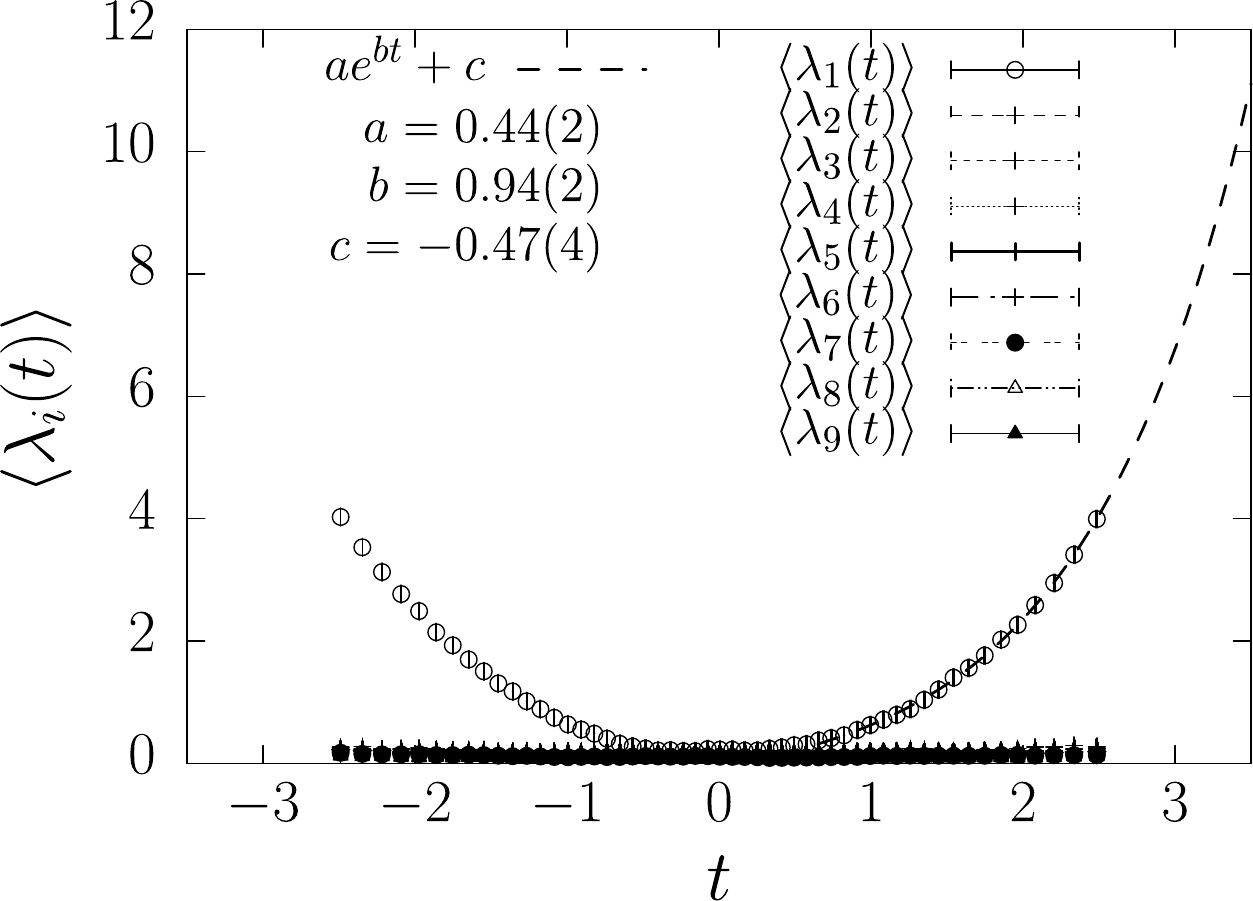}
    \includegraphics[width=0.33\hsize]{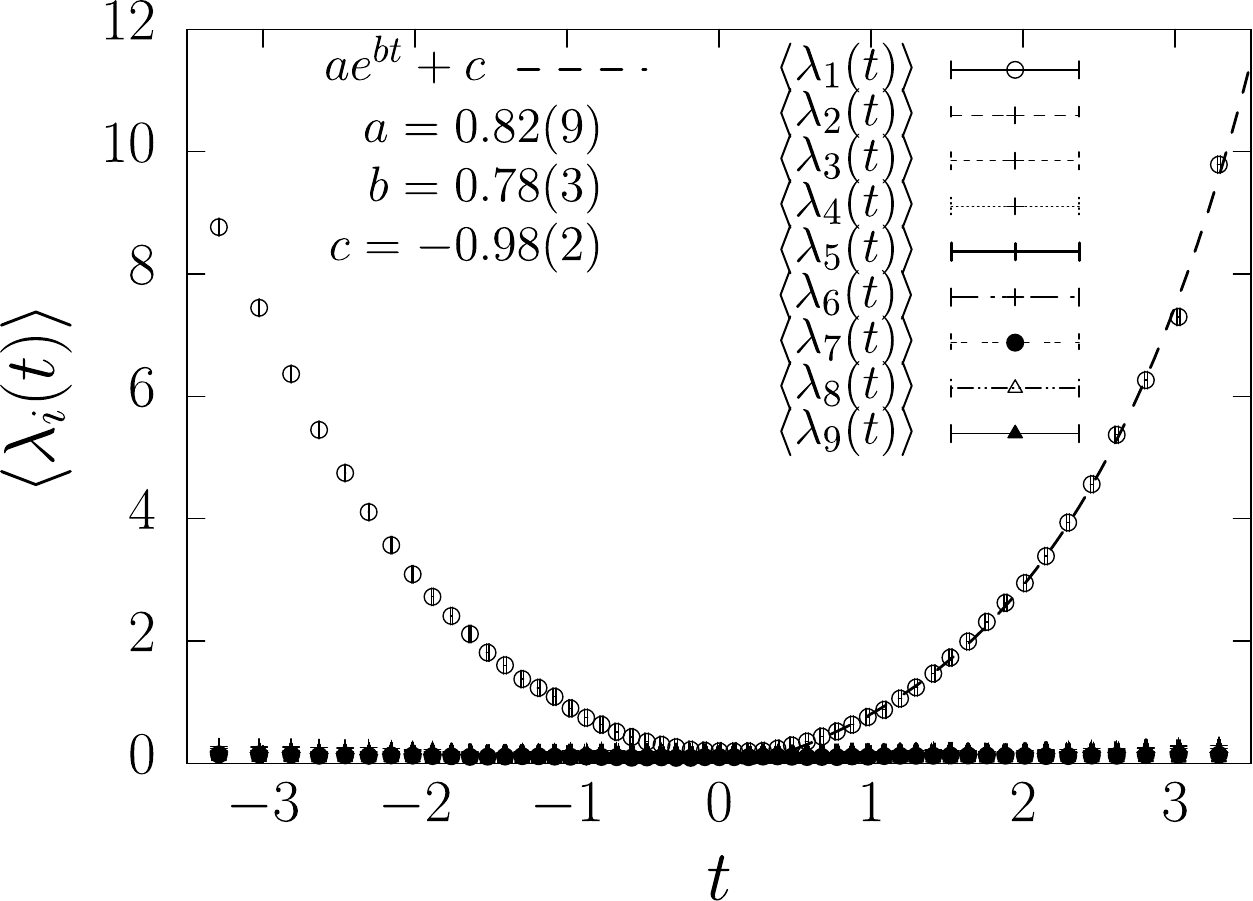}
    \includegraphics[width=0.33\hsize]{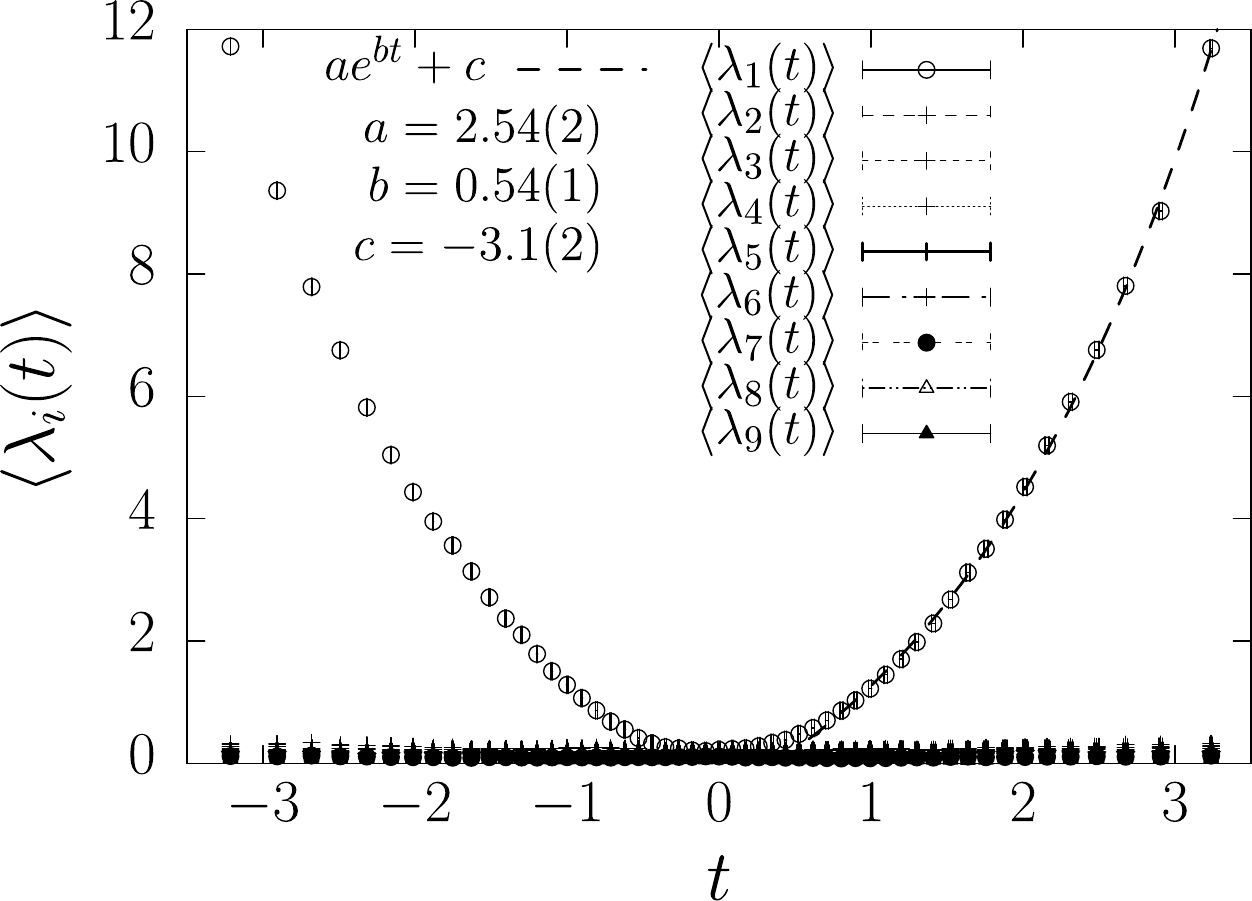}
    \caption{The eigenvalues of $T_{ij}(t)$ are plotted against time $t$. (Left Top) $\gamma=4.0$ and $m_{\rm f}=10$, (Right Top) $\gamma=2.6$ and $m_{\rm f}=10$, and (Bottom) $\gamma=2.6$ and $m_{\rm f}=5$. The dashed lines are obtained by exponential fits using $t \gtrsim 0.5$ data points.}
    \label{fig:gam_dep}
    \vspace{-1em}
\end{figure}

In the SO(9) symmetric case, the eigenvalues of $T_{ij}(t)$ are all equal in the large-$N$ limit \footnote{The small differences between them are a finite-$N$ effect.}, but in the SO(9) broken case they are not.
In Fig.\ref{fig:gam_dep}, we plot the eigenvalues of $T_{ij}(t)$ against the time $t$.
These figures show that the eigenvalues become equal around $t=0$.
On the other hand, we find that the SSB of SO(9) occurs at $t \sim 0.5$.
After the SSB, 1 out of the 9 eigenvalues grows exponentially.
In other words, 1-dimensional space grows exponentially.
By comparing Fig.\ref{fig:gam_dep} (Left Top) and (Right Top), we can see that the expansion becomes more pronounced as $\gamma$ decreases.
Furthermore, by comparing Fig.\ref{fig:gam_dep} (Right Top) and (Bottom), we see that the expansion becomes more pronounced as $m_{\rm f}$ decreases.

\vspace{-0.5em}
\section{Summary and discussion}
\vspace{-0.25em}
We studied the Lorentzian version of the type IIB matrix model numerically.
Since there is a strong sign problem, we used the complex Langevin method.
To avoid the singular drift problem we add a mass term for the fermionic matrices.
We denote this mass parameter by $m_{\rm f}$, and the original model is obtained after an $m_{\rm f} \to 0$ extrapolation.

The Lorentzian version of the model has been found to be equivalent to the Euclidean one under a contour deformation.
We confirm this point numerically for the fermion quenched model.
To obtain results inequivalent to the Euclidean model, we use the Lorentz invariant ``mass'' term which acts as an IR regulator.
We denote its mass parameter by $\gamma$.
We propose a novel large-$N$ limit, where one takes the $\gamma \to 0$ extrapolation after the large-$N$ limit is taken for specific $\gamma$.

Regarding the mass parameter $\gamma$, there are two phases depending on its value.
One appears at small $\gamma$, in which the model is found to be equivalent to the Euclidean model after a contour deformation.
The other phase appears at sufficiently large $\gamma$, in which time becomes real.
We call this phase the real time phase.

An important feature of the real time phase is that the spatial matrices have a band-diagonal structure.
This structure enables us to define block matrices which represent
the state of the universe at a given time.
We studied the time evolution of the universe using these block matrices.
We found that the real space appears at late times.
Moreover, a spontaneous symmetry breaking (SSB) of SO(9) occurs at some time, however, only 1-dimensional space expands exponentially after the SSB when the fermionic mass term is large ($m_{\rm f} \gtrsim 5$).

By focusing on the bosonic part of the action, the quantum fluctuations are suppressed when ${\rm Tr} [A_i, A_j] \sim 0$, which happens when only one spatial matrix is large.
This is the reason for the emergence of the 1-dimensional space at sufficiently large $m_{\rm f}$.
On the other hand, when $m_{\rm f} = 0$ and there are only two large matrices the Pfaffian becomes zero \cite{Krauth:1998xh,Nishimura:2000ds}.
Therefore, configurations in which only one or two of matrices are large
are strongly suppressed because the Pfaffian becomes small.
Thus, the emergence of expanding 1-dimensional space is suppressed by the presence of SUSY, and we expect the emergence of an expanding 3-dimensional space for sufficiently small $m_{\rm f}$.

\vspace{-0.5em}
\section*{Acknowledgements}
\vspace{-0.25em}
\setlength{\baselineskip}{13.5pt}
T.\;A., K.\;H. and A.\;T. were supported in part by Grant-in-Aid (Nos. 17K05425, 19J10002, and 18K03614, 21K03532, respectively)
from Japan Society for the Promotion of Science.  This research was supported by MEXT as ``Program for Promoting Researches on the
Supercomputer Fugaku'' (Simulation for basic science: from fundamental laws of particles to creation of nuclei, JPMXP1020200105) and JICFuS.
This work used computational resources of supercomputer Fugaku provided by the RIKEN Center for Computational Science (Project ID: hp210165, hp220174), and Oakbridge-CX provided by the University of Tokyo (Project IDs: hp200106, hp200130, hp210094, hp220074) through the HPCI System Research Project.
Numerical computations were also carried out on PC clusters in KEK Computing Research Center.  
This work was also supported by computational time granted by the Greek Research and Technology Network (GRNET) in the National HPC facility ARIS, under the project IDs SUSYMM and SUSYMM2.
K.\;N.\;A and S.\;K.\;P. were supported in part by a Program of Basic Research PEVE 2020 (No. 65228700) of the National Technical University of Athens.

\vspace{-0.5em}
\setlength{\baselineskip}{10.25pt}
\bibliographystyle{JHEP}
\bibliography{bib.bib}
\end{document}